\documentclass[letterpaper, 10 pt, conference]{ieeeconf}
\usepackage{cite}
\usepackage{amsmath,amssymb,amsfonts}
\usepackage{algorithmic}
\usepackage{graphicx}
\usepackage{textcomp}
\usepackage{xcolor}
\usepackage{float}
\usepackage{soul}

\IEEEoverridecommandlockouts                              

\title{\LARGE \bf
Trajectory Planning Using Tire Thermodynamics\\for Automated Drifting*\\
}

\author{Takao Kobayashi$^{1}$, Trey P. Weber$^{2}$ and J. Christian Gerdes$^{3}$
\thanks{$^{1}$Takao Kobayashi is with the Department of Mechanical Engineering,
        Stanford University, Stanford, CA 94305 USA
        {\tt\small tkoba@stanford.edu}}%
\thanks{$^{2}$Trey P. Weber is with the Department of Mechanical Engineering,
        Stanford University, Stanford, CA 94305 USA
        {\tt\small tpweber@stanford.edu}}%
\thanks{$^{3}$J. Christian Gerdes is with the Department of Mechanical Engineering,
        Stanford University, Stanford, CA 94305 USA
        {\tt\small cgerdes@stanford.edu}}%
}

\begin{document}
\maketitle
\thispagestyle{empty}
\pagestyle{empty}

\begin{abstract}
Automated vehicles need to estimate tire-road friction information,
as it plays a key role in safe trajectory planning and vehicle dynamics control.
Notably, friction is not solely dependent on road surface conditions, but also varies significantly depending on the tire temperature.
However, tire parameters such as the friction coefficient have been conventionally treated as constant values in automated vehicle motion planning.
This paper develops a simple thermodynamic model that captures tire friction temperature variation.  To verify the model, it is implemented into trajectory planning for automated drifting - a challenging application that requires leveraging an unstable, drifting equilibrium at the friction limits.
The proposed method which captures the hidden tire dynamics provides a dynamically feasible trajectory,
leading to more precise tracking during experiments with an LQR (Linear Quadratic Regulator) controller.
\end{abstract}


\section{INTRODUCTION}
Automated vehicles need to estimate tire-road friction information,
as it plays a key role in safe trajectory planning and vehicle dynamics control.
Notably, friction is not solely dependent on road surface conditions.  
The tire-road friction is known to exhibit significant temperature dependence owing to rubber characteristics and the internal structure\cite{t0}.
Recently, tire models capable of taking  tire temperature dependency and thermodynamics into account have been developed.
Mizuno \textit{et al.}\cite{t1} incorporated a tire surface temperature effect to the empirical ``Magic Formula''\cite{mf} tire model by approximating the tire thermodynamics as a first order delay system.
Sorniotti \cite{t2} enhanced an analytical tire model with temperature dependence and showed a significant difference in a vehicle motion when the temperature was allowed to vary.
For higher fidelity, detailed FEM (Finite Element Method) based models have also been developed\cite{t3}\cite{t4}.
Bosch \textit{et al.}\cite{t6} leveraged the latest Magic Formula model, which includes a tire thermodynamic model\cite{t7}, to predict the tire friction coefficient for an automated emergency braking simulation.  This work provides an excellent example of the application of modern tire models to vehicle control.

Tire thermodynamics can be applied not only to a dry road, but also to an icy road.
The quantity of water produced due to the heat generated by the frictional force at the tire contact patch substantially influences the friction\cite{t8}.
Leveraging the tire temperature dependence and the thermodynamics can be valuable information to estimate the tire friction, particularly in severe conditions.
However, there have been few examples incorporating the tire temperature dependence into actual control systems.
\begin{figure}[tbp]
\centerline{\includegraphics[width=1\linewidth]{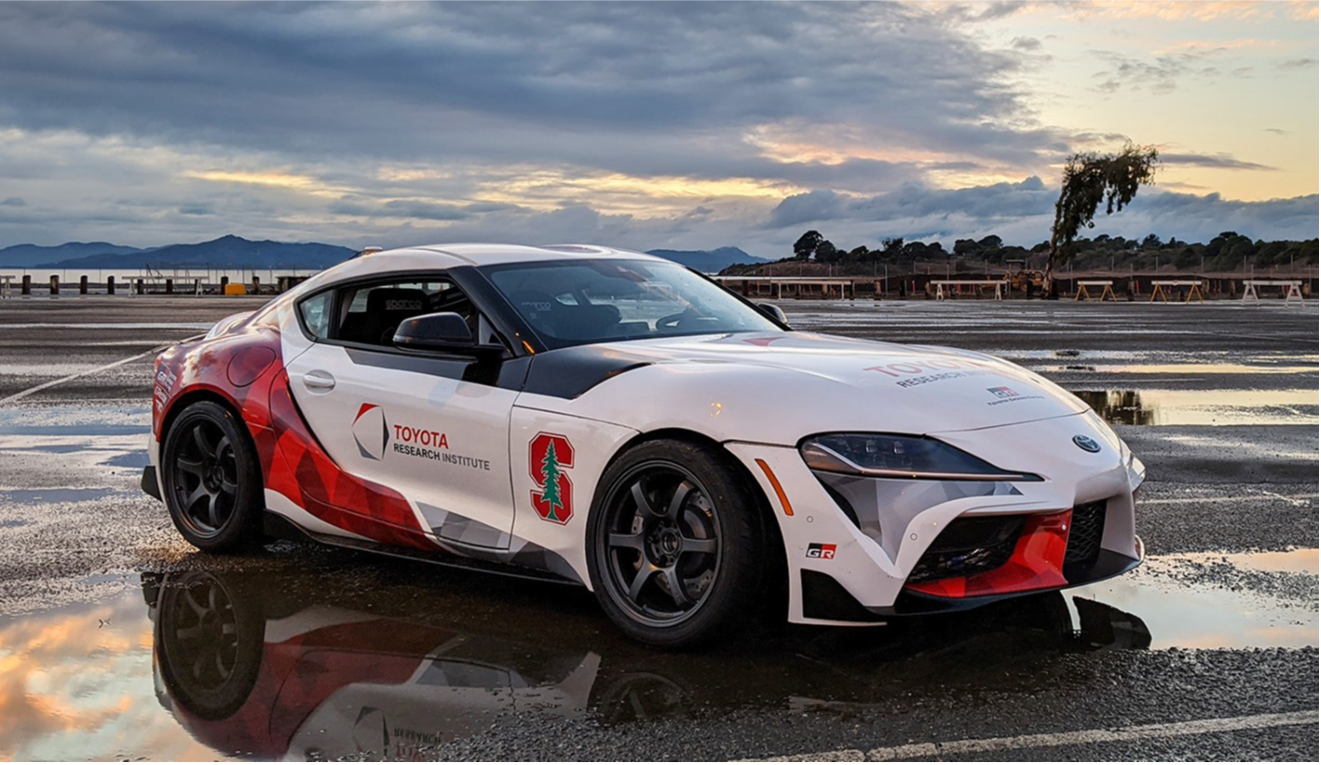}}
\caption{``Takumi'': An automated drifting platform}
\label{Takumi}
\end{figure}

In this paper, we address tire temperature dependence for automated vehicles. 
We choose automated drifting as a challenging application\cite{b1,b2} for such a model because a significant increase in the rear tire temperature and a large friction change have been observed through our experiments\cite{b3,b4}.
We develop a trajectory calculation method incorporating the tire thermodynamics that can handle drifting,
and demonstrate the effectiveness through experiments on a full-sized test vehicle (Takumi, shown in Fig.\ref{Takumi}). Essentially, the drifting equilibrium is unstable, but controllable\cite{ono}. 
For the trajectory calculation, a drifting equilibrium is generally computed from the assumption that the derivative of the vehicle dynamics state is zero. 
Since the drifting equilibrium changes gradually in accordance with the tire temperature even if the target velocity and slip angle are constant, the final trajectory is calculated as a set of the quasi-equilibria in which only the tire temperature changes.
A LQR controller is then used to track the planned trajectory\cite{b3}.
The gain is updated with the temperature change information along the path to stabilize the vehicle motion.
The longitudinal weight transfer and wheel speed dynamics are also taken into account to calculate more accurate traction\cite{b4}.

\section{Vehicle Modeling}
\subsection{Path Tracking Model}
The path tracking model in this paper is defined in a curvilinear coordinate system.
As shown in Fig. \ref{path}, the vehicle position is calculated relative to a desired reference path,
which is defined with the distance along the path \textit{s}.
The lateral tracking error \textit{e} is the distance from the path to the vehicle's center of mass.
And the angle error $\Delta \psi$ is between the vehicle's heading angle and the tangent of the reference path.
The path dynamics are described as below.

\begin{equation} 
\dot{e} = V_y \cos{\Delta \psi} + V_x \sin{\Delta \psi} \label{edot}
\end{equation}
\begin{equation} 
\dot{s} = \frac{V_x \cos{\Delta \psi} - V_y \sin{\Delta \psi}}{1-\kappa e} \label{sdot}
\end{equation}
\begin{equation} 
\dot{\Delta \psi} = r - \kappa \dot{s} \label{psidot}
\end{equation}
        
\subsection{Single Track Model}
\begin{figure}[t]
\centerline{\includegraphics[width=0.8\linewidth]{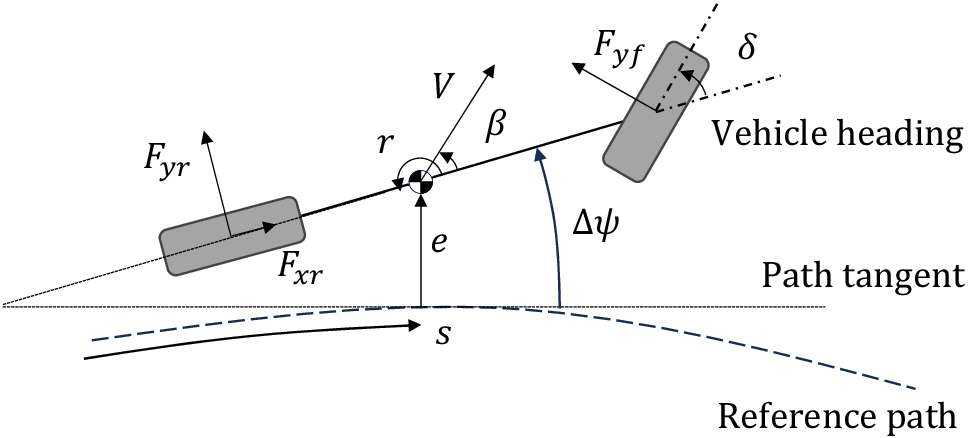}}
\caption{Single track model with reference path}
\label{path}
\end{figure}

We use the single track model which strikes a balance between model fidelity and simplicity as shown in Fig. \ref{path}.
$V_x(=V\cos\beta)$ and $V_y(=V\sin\beta)$ are the longitudinal and lateral velocities of the vehicle's center of mass, respectively.
$r$ is the yaw rate. We also include the wheel dynamics to the state equation\cite{b4}, where $\omega$ is the wheel speed of the rear tire.
The system inputs are the steering wheel angle $\delta$, the driving torque of the rear axle $\tau_r$, and the braking force of the front axle $F_{xf}$.

\begin{equation} 
\dot{V_x} = \frac{-F_{yf} \sin{\delta}+F_{xf}\cos{\delta}+F_{xr}}{m} + rV_y \label{ux}
\end{equation}
\begin{equation} 
\dot{V_y} = \frac{F_{yf} \cos{\delta}+F_{xf}\sin{\delta}+F_{yr}}{m} - rV_y \label{uy}
\end{equation}
\begin{equation} 
\dot{r} = \frac{aF_{yf} \cos{\delta}+aF_{xf}\sin{\delta}-bF_{yr}}{I_z} \label{r}
\end{equation}
\begin{equation} 
\dot{\omega} = \frac{\tau_{r} -R_eF_{xr}}{J}\label{wdot}
\end{equation}
Here, $F_{yf}$ and $F_{yr}$ are the lateral forces of the front and rear tires. 
Also, vehicle parameters are included above equations: $m$, mass, $I_z$, moment of inertia,
and $a$ and $b$, the distance from the center of mass to the front and rear axles, respectively.

Furthermore, we consider the longitudinal weight transfer $\Delta F_{z}$ to take account of the steer characteristics change due to the extremely high vehicle slip angle while drifting as below\cite{b4}. We approximate the dynamics as a first order delay system, where $K_z$ is the longitudinal weight transfer dynamics gain and $h_{cg}$ is the height of the center of gravity.
\begin{equation} 
\dot{\Delta F_{z}} = -K_{z}[\Delta F_{z}-\frac{h_{cg}}{L}(F_{xr} - F_{yf}\sin{\delta})] \label{Fz}
\end{equation}
Then, the vertical loads $F_{zf}$ and $F_{zr}$ of the front and rear axles are calculated as follows. 
\begin{equation} 
{F_{zf}} = \frac{bmg}{L}-\Delta F_{z} \label{Fzf}
\end{equation}
\begin{equation} 
{F_{zr}} = \frac{amg}{L}+\Delta F_{z} \label{Fzr}
\end{equation}
where, $g$ is the gravitational acceleration, and $L$ is the wheelbase($=a+b$).

\subsection{Tire Force Model}
For the front axle, the lateral force $F_{yf}$ is modeled using the Fiala brush tire model which assumes a pure lateral slip state, because it is not driven by the engine and rarely intervened by the brakes.
The front side slip angle is computed as follows.
\begin{equation} 
{\alpha_{f}} = \arctan\left({\frac{V_y + ar}{V_x}}\right)-\delta \label{af}
\end{equation}
Then, the front lateral force is obtained as follows.
\begin{equation} 
F_{yf} =
    \left\{
        \begin{aligned}
            &-C_\alpha \tan\alpha+\frac{C_\alpha^2}{3F_{ymax}}|\tan\alpha|\tan\alpha\\
            &-\frac{C_\alpha^3}{27F^2_{ymax}}\tan^3\alpha \;\;\; |\alpha| \leq \alpha_{slide}\\
            &-F_{ymax}sign(\alpha) \;\;\; |\alpha| < \alpha_{slide}
        \end{aligned}
    \right.
\label{Fyf}
\end{equation}
where, $C_\alpha$ is the cornering stiffness, $F_{ymax}$ is the maximum front lateral force ($=\mu F_z$).
and $\alpha_{slide} = \tan^{-1}{\left(\frac{3F_{ymax}}{C_\alpha}\right)}$.
The load dependency of the cornering stiffness is approximated as an affine function of the front load.
\begin{equation} 
C_\alpha = C_{\alpha1}F_{zf} + C_{\alpha0} \label{ca}
\end{equation}
On the other hand, the rear axle lateral force $F_{yr}$ is modeled using a combined slip model to represent the combined effect of the longitudinal force and lateral force, while drifting.
The rear side slip angle and slip ratio are computed as below.
\begin{equation} 
{\alpha_{r}} = \arctan\left({\frac{V_y - br}{V_x}}\right) \label{ar}
\end{equation}
\begin{equation} 
{\kappa_{r}} = \arctan\left({\frac{R_e\omega_r - V_x}{V_x}}\right) \label{kr}
\end{equation}
The respective tire forces are obtained as below.
\begin{equation} 
f = \sqrt{C_x^2\left({\frac{\kappa_r}{\kappa_r + 1}}\right) + C_y^2\left({\frac{\tan{\alpha_r}}{\kappa_r + 1}}\right)}
\label{f}
\end{equation}

\begin{equation}
F =
\begin{cases}
f-\frac{f^2}{3\mu_{r}F_{z}}+ \frac{f^3}{27(\mu_{r}F_{zr})^2}&\text{if $f \leq 3\mu_{r}F_{zr}$,}\\
\mu_{r}F_{zr}&\text{if $f > 3\mu_{r}F_{zr}$.}
\end{cases}
\end{equation}

\par
\par
\begin{equation} 
F_{xr} = \frac{FC_x\kappa_r}{f(\kappa_r+1)} \label{Fxr}
\end{equation}
\begin{equation} 
F_{yr} = \frac{FC_y\tan\alpha_r}{f(\kappa_r+1)} \label{Fyr}
\end{equation}
where $C_x$ and $C_y$ are the driving stiffness and cornering stiffness of the rear tire, respectively,
and $\mu_r$ is the friction coefficient of the rear tire.

\subsection{Tire Thermal Model}
\begin{figure}[tbp]
\centerline{\includegraphics[width=0.9\linewidth]{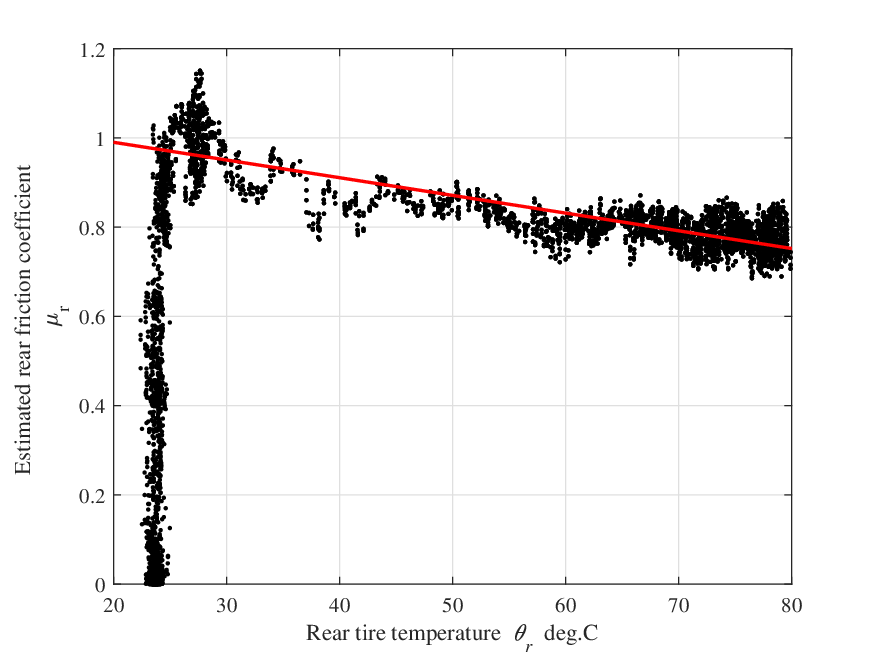}}
\caption{Tire friction map using tread temperature: the friction coefficient is estimated using an observer\cite{b4} and the parameters are identified from 30 deg.C, where the real tires are totally in the slip region. The points around 25 deg.C means the rear tires still have the margin from the friction limit, therefore it is impossible to estimate the friction parameters.}
\label{mumapMar}
\end{figure}

As described later, we have an access to tread surface temperature data 
and the friction coefficient is related to it in principle.
Here, the friction coefficient is approximated as an affine function of the rear tire temperature as shown in Fig.\ref{mumapMar}.
\begin{equation} 
\mu_r = \mu_{r1}\theta_{r} + \mu_{r0} \label{mur}
\end{equation}
The friction coefficient is calculated using an tire force observer\cite{b4}.
These parameters are identified to a test data while drifting, which means that the tire contact patch is operating in the sliding region completely.
The tread surface temperature has dynamics based on an equation of energy and heat transfer with the ambient environment.
\begin{equation} 
C_{tire}\dot{\theta_r} = Q_{tire} - KA_{tire}(\theta_r - \theta_{out}) \label{thetar}
\end{equation}
where, $C_{tire}$ is the heat capacity of the tire,
$KA_{tire}$ is the thermal conductance including the heat transfer with the ambient air
and the heat conduction with the contact patch.
And $Q_{tire}$ is the amount of heat generation due to the slip loss power\cite{takao} and the rolling resistance.
This heat generation can be calculated from the following equation.
\begin{equation} 
Q_{tire} = -\alpha_{tire}(V_{sx}F_{xr} + V_{sy}F_{yr}) - \epsilon_{tire}(F_zr)V_x \label{Qtire}
\end{equation}
where, $\alpha_{tire}$ is the partition coefficient which defines the fraction of the slip loss power added to the tire itself
(the rest is conducted to the road surface via the contact patch)\cite{t0},
$\epsilon_{tire}$ is the rolling resistance coefficient of the tire.
$V_{sx}$ and $V_{sy}$ are the longitudinal and lateral slip velocities, respectively, defined further below.
\begin{equation} 
V_{sx} = V_{x}\kappa_{r} \label{Vsx}
\end{equation}
\begin{equation} 
V_{sy} = \left(V_{y}-br\right) \sin{\alpha_{r}} \label{Vsy}
\end{equation}
While other mechanisms of tire heat transfer exist, such as oil evaporation from the tire while drifting, they are significantly more complicated to model
than the existing physical model which only considers convective heat transfer \cite{b1}.
Instead, we identify the dynamics as a lumped parameter system from an actual test for ease of the parameter identification, implementation and optimization stability.

\begin{figure}[tbp]
\centerline{\includegraphics[width=1.05\linewidth]{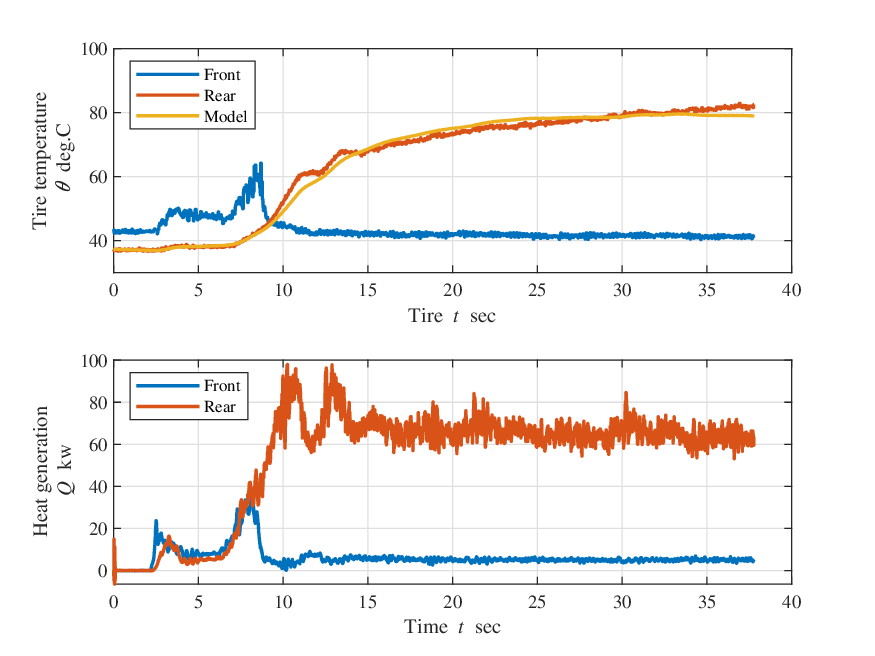}}
\caption{Tire thermodynamic model: Whereas the front tire temperature remains same, the rear tire temperature rises dramatically while drifting.}
\label{thermal}
\end{figure}

\begin{table}[htbp]
\caption{Vehicle \& tire parameters}
\begin{center}
\begin{tabular}{|c|c|c|}
\hline
\textbf{Parameter}& \textbf{Symbol}& \textbf{Value} \\
\hline
\text{Mass}& $M$&1496 \text{kg}\\
\text{Yaw moment of inertia}& $I_z$& 2241 kg$\cdot$m$^\text{2}$\\
\text{Distance from CG to front axle}& $a$&1.22 \text{m}\\
\text{Distance from CG to rear axle}& $b$&1.23 \text{m}\\
\text{Height of CG}& $h_{cg}$&0.45 \text{m}\\
$\Delta F_{z}$ \text{first-order dynamics gain}&$K_{z}$&5 1/s\\
\text{Drivetrain inertia}& $J$&15 kg$\cdot$m$^\text{2}$\\
\text{Tire radius}& $R_e$&0.32 m\\
\text{Front axle lateral parameters}& $C_{\alpha0}$&-18215 N/rad\\
& $C_{\alpha1}$&34.50 1/rad\\
\text{Rear axle longitudinal stiffness}& $C_{x}$&101E+3 N/rad\\
\text{Rear axle lateral stiffness}& $C_{y}$&103E+3 N/rad\\
\text{Rear axle friction parameters}& $\mu_{r0}$& 1.070\\
& $\mu_{r1}$& -3.967E-3 1/deg.C\\
\text{Rear tire heat capacity}& $C_{tire}$& 4.905 kJ/K\\
\text{Rear tire thermal conductance}& $KA_{tire}$& 0.762 kW/K\\
\text{Rear tire partition coefficient}& $\alpha_{tire}$& 0.5\\
\text{Rear tire rolling resistance coefficient}& $\epsilon_{tire}$& 0.01\\
\hline
\end{tabular}
\label{tab1}
\end{center}
\end{table}

\section{Trajectory Planning and Control}
We generate reference trajectories with and without the tire thermal model to examine the effectiveness of the proposed model. Here, two different types of trajectory are calculated. One is a steady state drifting where the drifting equilibrium seems consistent along the path, another is a figure 8 course as a dynamic drifting, including a transient maneuver. Also we reflect the system dynamic characteristics change due to the tire temperature in designing a LQR feedback controller.
\subsection{Steady state drifting trajectory}
\begin{figure}[htbp]
\centerline{\includegraphics[width=0.75\linewidth]{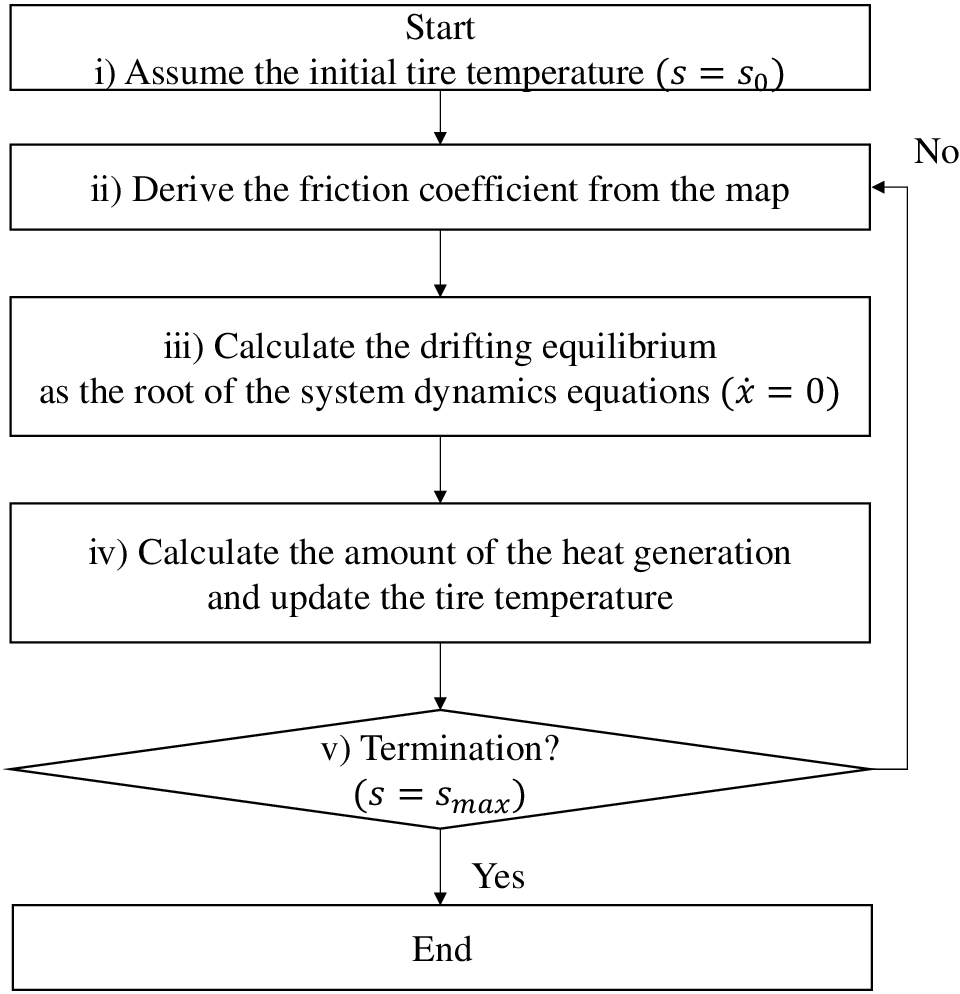}}
\caption{Flowchart to compute a set of drifting equilibria using the tire thermodynamics model}
\label{equilibria}
\end{figure}

Even if the turning radius and the side slip angle are same, 
the equilibrium point is different when the road friction changes due to the tire temperature.
Although it is not the steady state anymore strictly speaking, we assume the reference trajectory as a set of quasi-steady state drifting.
From the vehicle dynamics equations in the last section,
the state vector is $x_s=[r, V, \beta, \omega_r, \Delta F_{zlong}]^T$ and the input vector is $u_s=[\delta, F_{xf}, \tau_r]^T$.
Fig.\ref{equilibria} shows the flowchart to compute the drifting equilibria.
At first, we assume the initial tire temperature.
Then, the drifting equilibrium can be obtained finding the root of the system equations.
Based on the solution, we compute the amount of the tire heat generation and update the tire temperature at the next step. 

\subsection{Dynamic drifting trajectory}
We also validate the effectiveness of the thermal model in a figure 8 course as a more dynamic trajectory.
A transient maneuver is added between steady state drifting.
Since the solution is not static anymore, we solve the reference motion as a optimization problem.
Here, the state vector is
$x_d=[r, V, \beta, \psi, \omega_r, \Delta F_{zlong}, X, Y, \delta, \tau, \theta_r]^T$
and the input vector is $u_d=[\dot{\delta}, \dot{F_{xf}}, \dot{\tau_r}]^T$.
The input vector consists of slew rates of the steering angle, front braking force and rear wheel torque. These slew rates are costed to avoid harshness. The transition distance can be tuned shorter if results in a more dynamic change.
Thus, the cost function is as below:
\begin{equation} 
J = \sum_{k=1}^{N}({u_{dk}Ru_{dk}})+s_{N}^Tk_{s}s_{N}
\label{J}
\end{equation}
\begin{equation} 
R = diag([k_{\dot{\delta}}, k_{\dot{F}_{xf}}, k_{\dot{\tau}_r}])
\label{Q}
\end{equation}
The braking force is not used here, so its cost weight is set to zero ($k_{\dot{F}_{xf}} = 0$).
By increasing the value of $k_s$, the vehicle changes the direction more dynamically.
The next path information is given including the curvature ($\kappa_{final}$),
and the lateral position of the center of the two circular paths are set to same for the experiments.
Finally, the optimization problem including the constraints is as below:
\begin{equation}
\begin{aligned}
\underset{x, u}{\text{minimize}}\quad J\\
\text{subject to}\quad x_1 &= x_{initial}\\
r_{N}/V_{N} &= \kappa_{final}\\
\beta_{N} &= \beta_{final}\\
\dot{u_N} &= 0\\
\delta_{min} &\leq \delta_k \leq \delta_{max}\\
\dot{\delta}_{min} &\leq \dot{\delta_k} \leq \dot{\delta}_{max}\\
\tau_{min} &\leq \tau_k \leq \tau_{max}\\
\dot{\tau}_{min} &\leq \dot{\tau_k} \leq \dot{\tau}_{max}
\label{optimizationproblem}
\end{aligned}
\end{equation}
Here, the state is numerically integrated with Runge-Kutta method (RK4).
The problem is discretized with 100 steps ($N=100$).

\subsection{Controller}
We use LQR as a controller to track the reference trajectory\cite{b3}.
The linearized state vector is $x = [V_x - \hat{V}_x, V_y - \hat{V}_y, r - \hat{r}, \omega - \hat{\omega}, \Delta\psi  - \Delta\hat{\psi}, e-\hat{e}]^T$,
and the input vector is $u = [\delta - \hat{\delta}, F_{xf} - \hat{F}_{xf}, \tau - \hat{\tau}]$.
Note that $\hat{ }$ means the reference value at the equilibrium from the trajectory generation.
The state equation can be obtained from the numerical derivatives of the above equations.
We can analytically differentiate only $\Delta \psi$ and $e$ with respect to the state vector $x$.
By substituting the reference values ($\hat{e}=0$), the state equation can be expressed as follows.

\begin{equation*} 
\dot{x} = Ax+Bu
\label{ss1}
\end{equation*}

\footnotesize
\begin{equation*} 
A=\\
\begin{bmatrix}
\frac{\partial\dot{V_x}}{\partial V_x}&\frac{\partial\dot{V_x}}{\partial V_y}&\frac{\partial\dot{V_x}}{\partial r}&\frac{\partial\dot{V_x}}{\partial \omega}&0&0\\
\frac{\partial\dot{V_y}}{\partial V_x}&\frac{\partial\dot{V_y}}{\partial V_y}&\frac{\partial\dot{V_y}}{\partial r}&\frac{\partial\dot{V_y}}{\partial \omega}&0&0\\
\frac{\partial\dot{r}}{\partial V_x}&\frac{\partial\dot{r}}{\partial V_y}&\frac{\partial\dot{r}}{\partial r}&\frac{\partial\dot{r}}{\partial \omega}&0&0\\
\frac{\partial\dot{\omega}}{\partial V_x}&\frac{\partial\dot{\omega}}{\partial V_y}&\frac{\partial\dot{\omega}}{\partial r}&\frac{\partial\dot{\omega}}{\partial \omega}&0&0\\
-\hat{\kappa}\hat{C}&\hat{\kappa}\hat{S}&1&0&\hat{\kappa}(\hat{V_x}\hat{S}+\hat{V_y}\hat{C})&-\hat{\kappa}^2(\hat{V_x}\hat{C}-\hat{V_y}\hat{S})\\
\hat{S} & \hat{C} &0 &0 &-\hat{V_y} \hat{S}+\hat{V_x} \hat{C} &0
\end{bmatrix}
\label{ss2}
\end{equation*}
\normalsize

\begin{equation} 
B= 
\begin{bmatrix}
\frac{\partial\dot{V_x}}{\partial \delta}&\frac{\partial\dot{V_x}}{\partial F_{xf}}&\frac{\partial\dot{V_x}}{\partial \tau}\\
\frac{\partial\dot{V_y}}{\partial \delta}&\frac{\partial\dot{V_y}}{\partial F_{xf}}&\frac{\partial\dot{V_y}}{\partial \tau}\\
\frac{\partial\dot{r}}{\partial \delta}&\frac{\partial\dot{r}}{\partial F_{xf}}&\frac{\partial\dot{r}}{\partial \tau}\\
\frac{\partial\dot{\omega}}{\partial \delta}&\frac{\partial\dot{\omega}}{\partial F_{xf}}&\frac{\partial\dot{\omega}}{\partial \tau}\\
0&0&0\\
0&0&0
\end{bmatrix}
\label{ss3}
\end{equation}
where, $\hat{C}=\cos\hat{\psi}$ and $\hat{S}=\sin\hat{\psi}$.
Then, the control input is computed as $u = -Kx$. Here, $K$ is the gain matrix from the LQR problem.
The cost matrices $Q_{c}$ and $R_{c}$ are as below.
\begin{equation} 
\begin{split}
Q_{c} &= diag(k_{V_x}, k_{V_y}, k_{r}, k_{\omega}, k_{\Delta \psi},k_{e})\\
R_{c} &= diag(k_{\delta}, k_{F_{xf}}, k_{\tau})
\label{QR}
\end{split}
\end{equation}
Note that 
The system matrices change due to the tire temperature along the trajectory even with the constant radius.
Therefore, we update the gain matrix at each point of 0.25m.
The obtained gain matrix data is stored along the trajectory, and then the gain matrix with the nearest $s$ is chosen.

\begin{table}[tb]
\caption{Trajectory planning \& Control parameters}
\begin{center}
\begin{tabular}{|c|c|c|}
\hline
\textbf{Parameter}& \textbf{Symbol}& \textbf{Value} \\
\hline
\text{Cost on steering slew rate}& $k_{\dot{\delta}}$& 1E+4 (rad/s)$^{-2}$\\
\text{Cost on rear wheel torque slew rate}& $k_{\dot{\tau_r}}$&100 (kN$\cdot$m/s)$^{-2}$\\
\text{Cost on transition distance}& $k_{s}$&200 m$^{-2}$\\
\text{Maximum steering angle}& $\delta_{max}$& 43 deg\\
\text{Minimum steering angle}& $\delta_{min}$& -43 deg\\
\text{Maximum steering angle rate}& $\dot{\delta}_{max}$& 90 deg/s\\
\text{Minimum steering angle rate}& $\dot{\delta}_{min}$& -90 deg/s\\
\text{Maximum torque}& $\tau_{max}$& 3.5 kN$\cdot$m\\
\text{Minimum torque}& $\tau_{min}$& -1 kN$\cdot$m\\
\text{Maximum torque rate}& $\dot{\tau}_{max}$& 2 kN$\cdot$m/s\\
\text{Minimum torque rate}& $\dot{\tau}_{min}$& -4 kN$\cdot$m/s\\
\text{Cost on longitudinal velocity}& $k_{V_x}$& (0.5 m/s)$^{-2}$\\
\text{Cost on lateral velocity}& $k_{V_y}$&(1 m/s)$^{-2}$\\
\text{Cost on yaw rate}& $k_{r}$&(0.6 rad/s)$^{-2}$\\
\text{Cost on wheel speed}& $k_{\omega}$&(10 rad/s)$^{-2}$\\
\text{Cost on heading angle}& $k_{\Delta \psi}$&(10 deg)$^{-2}$\\
\text{Cost on lateral error}& $k_{e}$&(0.2 m)$^{-2}$\\
\text{Cost on steering angle}& $k_{\delta}$&(2 deg)$^{-2}$\\
\text{Cost on front braking force}& $k_{F_{xf}}$&(500 N)$^{-2}$\\
\text{Cost on rear wheel torque}& $k_{\tau}$&(500 N$\cdot$m)$^{-2}$\\
\hline
\end{tabular}
\label{tab2}
\end{center}
\end{table}

\section{Experimental Validation}
We validate the effectiveness of the thermodynamic model by comparing the tracking performance to constant friction models experimentally.

\subsection{Test Platform}
Takumi is as an autonomous drifting platform modified heavily based on a Toyota Mk-V Supra (A90).
The maximum power and torque is 526 HP and 536 ft-lbs, respectively.
An electric motor and encoder have been installed in the steering system to control the road wheel angle. Takumi also has a brake-by-wire and throttle-by-wire so that these actuators can be controlled by the tracking controller via a dSPACE Micro Autobox I\hspace{-1.2pt}I.
An Oxford Technical Systems RT4003 GNSS/INS with dual GPS antennae measures the state information described above and sends them to the control system.
Also, each wheelhouse has an infrared-type temperature sensor to measure the tire tread surface temperature.
The sensor has 3 channels in the tire width direction, but we average the output data to capture the friction change due to the temperature relatively.

\subsection{Steady state drifting}
\begin{figure}[H]
\centerline{\includegraphics[width=1.05\linewidth]{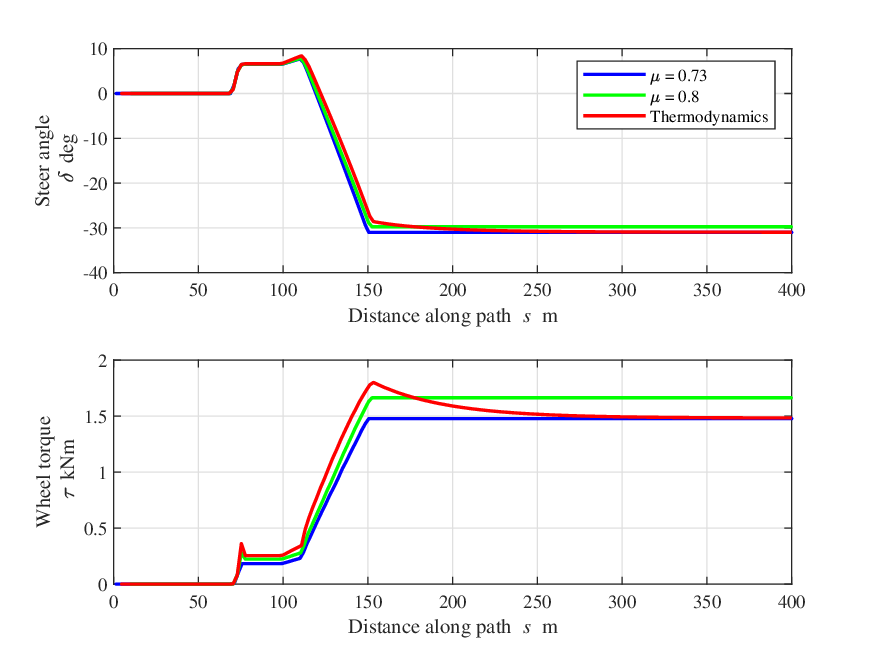}}
\caption{Steering angle and wheel torque commands for steady state drifting: The thermodynamic model changes the inputs in accordance with the rear tire friction changes.}
\label{sssim}
\end{figure}

The target path is: the turning radius at 15 meter, the side slip angle at -40 deg (counter-clockwise).
As a comparison, we test trajectories generated with constant friction coefficients.
The constant friction coefficients are set to 0.73 and 0.8, respectively. 
The generated trajectory information is shown in Fig.\ref{sssim}.
Whereas the inputs remain same while after the initiation of drifting for the cases of the constant friction,
they change along the trajectory when the tire thermodynamics are taken into account.

\begin{figure}[H]
    \centering
    \includegraphics[width=1.05\linewidth]{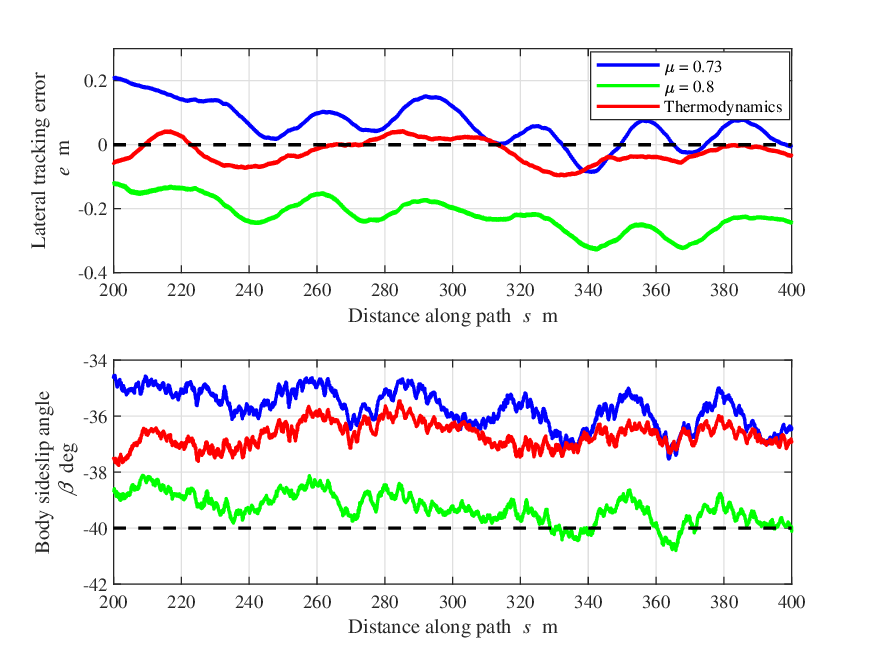}
    \caption{Steady state tracking error: the tracking error of the tire thermodynamic model is consistent along the path and kept within 10 cm, while Takumi is forced to go outside($e$ becomes negative for the other two trajectories.)}
    \label{sse}
\end{figure}

\begin{figure}[H]
\centerline{\includegraphics[width=1.05\linewidth]{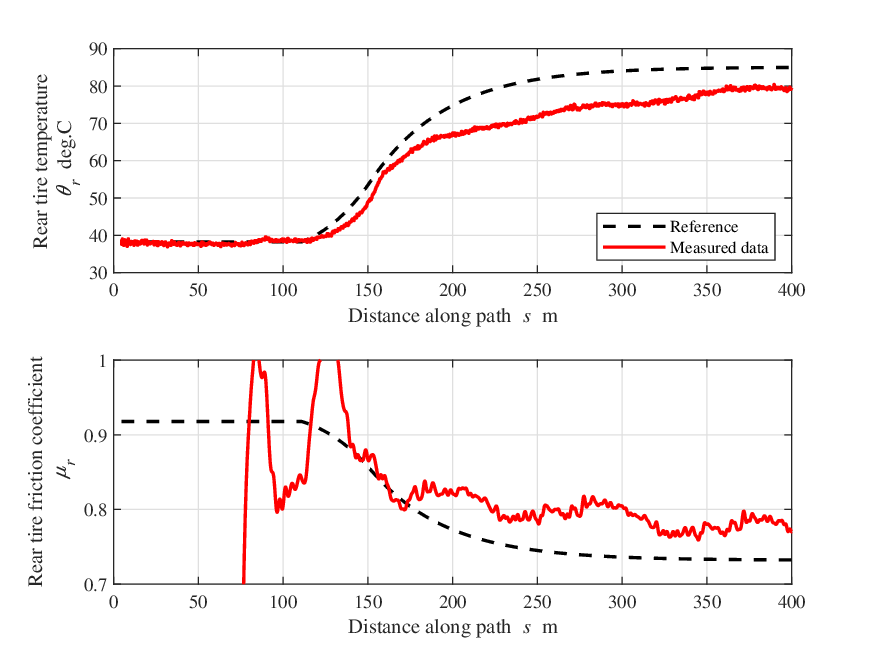}}
\caption{Comparison between the generated trajectory and actual test in the rear tire temperature and road friction: as a reference, the preplanned temperature and friction before the steady state drifting are depicted.}
\label{sst}
\end{figure}

Fig. \ref{sse} shows the tracking error of each case. The proposed method minimizes and stabilizes the tracking error within 10 cm.
Takumi is forced to go outside ($e$ becomes negative) for the trajectories with the constant friction values,
because the rear tires are gradually losing grip while drifting.
Still the proposed method has a side slip error to some extent, but the error is within 10 \% of the target angle (40 deg) and is kept constant as well as the tracking error.
Fig.\ref{sst} compares the actual tire temperature and the friction coefficient with the pre-planned ones.
There is an error in the thermal model: the model overestimates the temperature rise, leading to underestimation of the road friction.
But it captures the gradual change in the friction while drifting.

\begin{figure}[h]
\centerline{\includegraphics[width=0.85\linewidth]{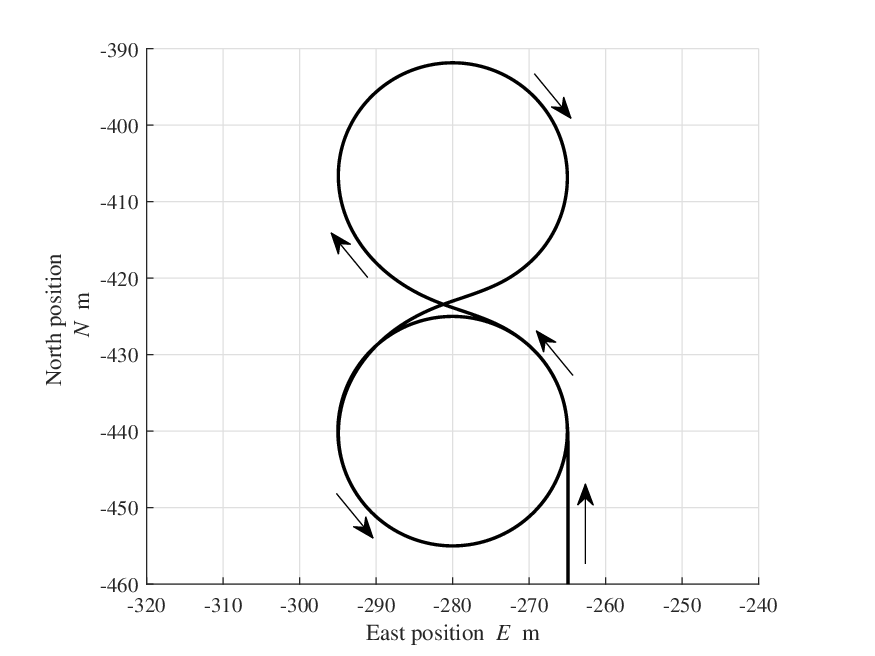}}
\caption{Preplanned reference trajectory generated from tire thermodynamic model}
\label{fig8}
\end{figure}
\subsection{Dynamic drifting}
\begin{figure}[t]
\centerline{\includegraphics[width=1.05\linewidth]{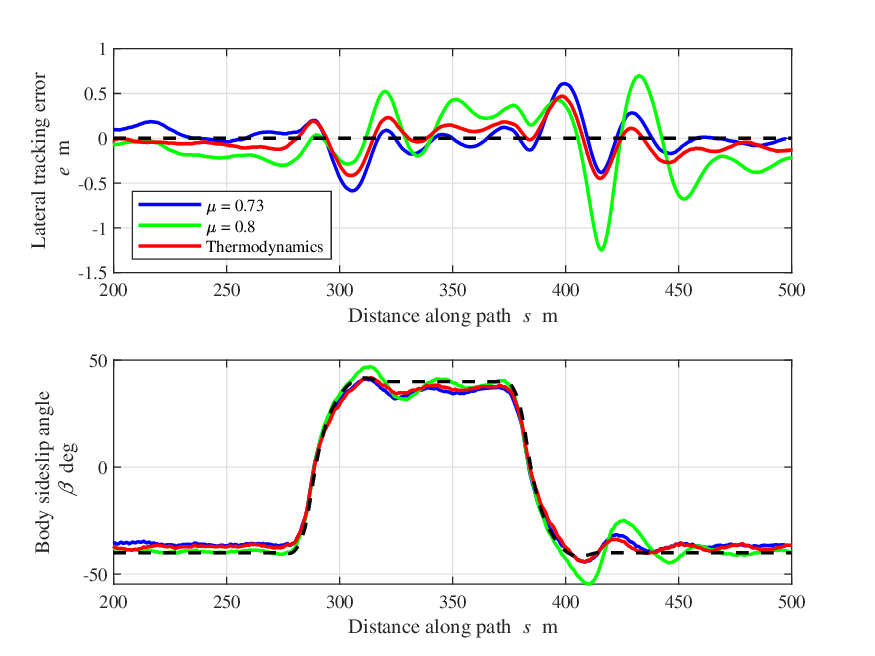}}
\caption{Tracking error in figure 8: the tracking error of the tire thermodynamic model is symmetrical along the path centerline, which means the model captures the friction change during the task.}
\label{f8e}
\end{figure}

\begin{figure}[t]
\centerline{\includegraphics[width=1.05\linewidth]{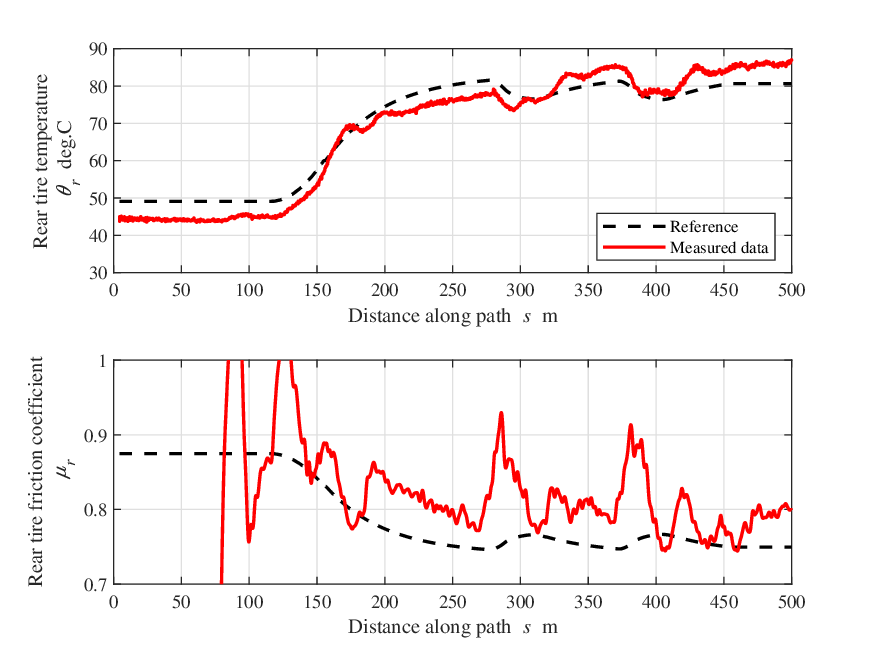}}
\caption{Comparison between the generated trajectory and actual test in the rear tire temperature and road friction in figure 8: although the actual data has spikes in the friction, the pre-planned friction captures the change relatively.}
\label{f8t}
\end{figure}

As a dynamic drifting, Takumi drifts a figure 8 course with the side slip angle at 40 deg. The two circles of 15 meters radius are located next to each other as shown in Fig.\ref{fig8}.
Fig. \ref{f8e} shows the tracking error of each case. 
At the transitions, ($s$ = 280 - 320 m and 370 - 410 m), the lateral tracking error is totally minimized and stable with the proposed method (within 50 cm).
The preplanned tire temperature and friction are compared with the test data in Fig.\ref{f8t}.
At the transitions, the tire begins to operate at the adhesive region, because Takumi is required to change the turning direction rapidly.
As a result, the temperature drops slightly and the friction increases.
Around $s$ = 300 m, Takumi goes inside ($e$ becomes negative) when the friction is assumed to be 0.73, because the friction is higher than planned.
At this point the constant friction 0.8 performs well, but it has the largest error and less damped motion at the second transition around $s$ = 400 m, because the friction is not as high as planned.
Overestimating friction causes the worst behavior.
Notably, the shape of the tracking error is symmetrical, which means that the friction change is captured well relatively. 
There are spikes of the friction at the transitions. The reason is considered as the difference in the tire friction between the adhesive region and sliding region in the contact patch.


\section{Discussion}
\begin{figure}[t]
\centerline{\includegraphics[width=1\linewidth]{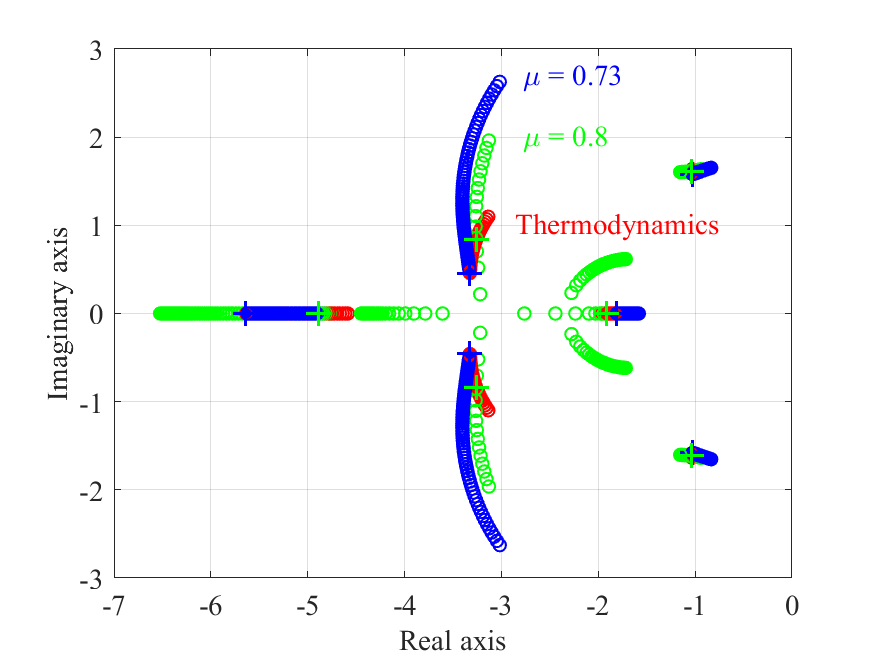}}
\caption{Pole placement: + indicates the original pole when the friction is assumed to be constant. The closed-loop dynamics of the thermodynamics model is almost consistent because the gain matrix can compensate the system matrix change due to the temperature change.}
\label{pp}
\end{figure}

As shown in Fig.\ref{sse} and Fig.\ref{f8t},
the tracking errors of the proposed method look consistent from a macroscopic view.
The trajectories based on the constant friction are not feasible,
because Takumi loses the traction to track them.
Up close, the proposed method seems to damp the tracking errors, too.
Fig.\ref{pp} compares the pole placement of the closed loop system ($\dot{x} = (A-BK)x$) during steady state drifting between the three trajectories.
The actual system has the temperature dependency. Therefore the friction change is taken into account of when calculating the system matrices $A, B$.
The gain matrix $K$ is consistent for the case of the constant friction, because the friction is assumed to be constant when calculating the LQR gain.
As a result, the closed-loop poles change along the path.
In contrast, the poles of the thermodynamics model are more consistent.
Because the gain matrix of the thermodynamics model can compensate the system matrices change.
Notably, the case of the friction 0.8 changes dramatically, because the actual friction is lower than that.
The proposed method contributes to the robustness of the feedback control system design.
Also it can be applied to enhance a nonlinear model predictive control capable of taking account of constraints of actuators explicitly and updating the system dynamics\cite{b4}.

When it comes to the tire temperature, there is room to improve the prediction accuracy.
The ambient air temperature and track surface temperature change even in one day.
Moreover, the radiation heat directly affects the tire temperature potentially.
As a result, the pre-planned friction also has an error to some extent.
It is also affected by the tire wear which means a long term degradation. 
Despite these complexities, a simple model of tire thermodynamics can improve the performance of state-of-the-art trajectory planning and control systems in extreme driving situations.
A data driven method may be able to capture this complicated physical phenomenon which is crucial for vehicle dynamics.
\section{CONCLUSION}
Tire performance changes with the temperature. However, in the vehicle control literature these parameters have traditionally been treated as constant values.
This paper demonstrates the implementation of the tire thermodynamic model into the trajectory planning for automated drifting.
The experimental result shows the proposed method minimizes and stabilizes the lateral tracking error,  
because it provides the dynamically feasible trajectory with less model mismatching error and ensures the more appropriate controller design based on the predicted tire temperature.

\addtolength{\textheight}{-12cm}   


\section*{ACKNOWLEDGMENT}
The authors also thank Toyota Research Institute and Bridgestone Corporation
for their generous support of this work.

\end{document}